\documentstyle[sprocl]{article}

\def\Journal#1#2#3#4{{#1} {\bf #2}, #3 (#4)}

\def\NPB{{\em Nucl. Phys.} B}
\def\PLB{{\em Phys. Lett.}  B}

\def\PRD{{\em Phys. Rev.} D}
\def\ZPC{{\em Z. Phys.} C}

\def\be{\begin{equation}}
\def\ee{\end{equation}}
\def\bea{\begin{eqnarray}}
\def\eea{\end{eqnarray}}

\begin{document}

\title{RESUMMATION OF LARGE LOGARITHMS IN HE CHARM STRUCTURE FUNCTION} 

\author{W.L. van Neerven \footnote{Talk presented at the DIS-98 Workshop,
Brussels, Belgium, April 4-8, 1998.}}

\address{DESY-Zeuthen, Platanenallee 6, D-15738 Zeuthen, Germany
\footnote{On leave of absence from Instituut-Lorentz, University of Leiden
, P.O. Box 9506, 2300 RA Leiden, The Netherlands.}}


\maketitle\abstracts{ 
We show how one can resum the large logarithms which occur in the 
charm component of the structure function $F_{k,c}(x,Q^2,m_c^2)$ ($k=2,L$)
taken at three light flavours, in the limit $Q^2 \gg m_c^2$. This
resummation leads to an expression of the above quantity for four light
flavours including the charm quark density. The latter is related to 
the parton densities in the three light flavour number scheme.}

The study of charm production in deep-inelastic electron-proton scattering
has become an important issue in the extraction of parton densities in the
proton. The reason is that in the kinematical region covered by the
HERA experiments \cite{der} \cite{adl} the charm component of the structure  
function $F_{2,c}(x,Q^2,m_c^2)$ constitutes about 25 $\%$ 
of the total structure function $F_2(x,Q^2)$ . Therefore
the analysis of the data to yield parton densities can no longer treat charm
electroproduction as a small correction. In a three flavour number scheme
(TFNS) the charm component of the structure function 
can be written as convolutions, denoted by the symbol $\otimes$, of parton 
densities and heavy quark coefficient functions. For $k=2,L$ it is given by
\bea
\label{eqn:1}
&& F_{k,c}^{\rm EXACT}(n_f,x,Q^2,m_c^2) = \frac{1}{n_f} \sum_{i=1}^{n_f} e_i^2
\Big [\Sigma\Big(n_f,\mu^2\Big) \otimes 
L_{k,q}^{\rm S}\Big(n_f,\frac{Q^2}{m_c^2},\frac{m_c^2}{\mu^2}\Big)
\nonumber\\[1ex]
&& + G\Big(n_f,\mu^2\Big) \otimes
L_{k,g}^{\rm S}\Big(n_f,\frac{Q^2}{m_c^2},\frac{m_c^2}{\mu^2}\Big)
+ n_f \Delta\Big(n_f,\mu^2\Big) \otimes
L_{k,q}^{\rm NS}\Big(n_f,\frac{Q^2}{m_c^2}, \frac{m_c^2}{\mu^2}\Big) \Big ]
\nonumber\\[1ex]
&& e_c^2 \Big [ \Sigma\Big(n_f,\mu^2\Big) \otimes
H_{k,q}^{\rm S}\Big(n_f,\frac{Q^2}{m_c^2},\frac{m_c^2}{\mu^2}\Big)
+  G\Big(n_f,\mu^2\Big) \otimes
H_{k,g}^{\rm S}\Big(n_f,\frac{Q^2}{m_c^2},\frac{m_c^2}{\mu^2}\Big)
\Big ] 
\nonumber\\
\,,
\eea
where $e_c$ stands for the charge of the charm quark and the number of light 
flavours is three ($n_f=3$).
The functions $\Sigma$ and $\Delta$ stand for the singlet (S) and non-singlet 
(NS) combinations of parton densities respectively and $G$ represents
the density of the gluon.
$L_{k,i}$ and $H_{k,i} ( i=q,g )$ represent the heavy quark coefficient
functions which can also be separated into singlet and
non-singlet parts. The functions $L_{k,i}$
originate from the reactions where the virtual photon couples to the light
quarks (u, d, s, ${\bar {\rm u}}$, ${\bar {\rm d}}$ and ${\bar {\rm s}}$),
whereas the $H_{k,i}$ describe the reactions
where the virtual photon is attached to the c ($\bar {\rm c}$) quark.
Finally we have included in Eq. (\ref{eqn:1}) that part of the light parton 
component of the proton structure function containing charm quark loop
contributions to the light parton (u,d,s,g) coefficient functions only.
Up to next-to-leading order (NLO)
the exact form of the heavy quark coefficient functions
can be found in \cite{lrsn}, \cite{rsn}. If we take the limit $Q^2\gg m_c^2$
these functions behave like
\bea
\label{eqn:2}
H_{k,i}^{{\rm ASYMP},(l)}(z,\frac{Q^2}{m_c^2},\frac{m_c^2}{\mu^2})
=\sum_{l=1}^{\infty} 
\Big (\frac{\alpha_s}{\pi}\Big )^l \sum_{n+m\leq l} a_{k,i}^{nm}(z)
\ln^n\Big(\frac{m_c^2}{\mu^2}\Big)\ln^m\Big(\frac{Q^2}{m_c^2}\Big)\,,
\eea
with an analogous expression for $L_{k,i}^{\rm ASYMP}$. The asymptotic
heavy quark coefficient functions have been calculated in \cite{bmsmn}.
When the latter are substituted in Eq. (\ref{eqn:1}) we obtain the asymptotic
charm component of the structure function which is equal to
\bea
\label{eqn:3}
F_{k,c}^{\rm ASYMP}(3,x,Q^2, m_c^2 )= \lim_{Q^2 \gg m_c^2}
\Big[F_{k,c}^{\rm EXACT}(3,x,Q^2,m_c^2)\Big] \,.
\eea
A NLO analysis \cite{bmsn1}
reveals that for $Q^2 > 20 ({\rm GeV/c})^2$ and $x < 0.01$, 
$F_{2,c}^{\rm ASYMP}$ coincides with $F_{2,c}^{\rm EXACT}$ (\ref{eqn:1}). 
Notice that hese values are not that far above threshold. However in the case
of the longitudinal structure function $F_{L,c}$ the value of $Q^2$ is
much larger and it amounts to $1000~({\rm GeV/c})^2$ for $x < 0.01$. 
The above findings
show that in the kinematical regime covered by the HERA experiments the
large logarithms in Eq. (\ref{eqn:2}) entirely determine $F_{2,c}$ but not
$F_{L,c}$. These large corrections vitiate the perturbation series so that
they have to be resummed in all orders of perturbation theory. This 
procedure has been carried out in \cite{bmsn1} and can be described as follows.
First we add the light parton (u,d,s,g) component of the proton structure
function denoted by $F_k(3,x,Q^2)$ to $F_{k,c}^{\rm ASYMP}$ (\ref{eqn:3}).
Then we perform mass factorization on the asymptotic heavy quark coefficient
functions. For $H_{k,i}^{\rm ASYMP}$ the mass factorization can be roughly
written as
\bea
\label{eqn:4}
H_{k,i}^{\rm ASYMP}\Big (3,\frac{Q^2}{m_c^2},\frac{m_c^2}{\mu^2}\Big )
=A_{ji}\Big(3,\frac{m_c^2}{\mu^2}\Big)
\otimes {\cal C}_{k,j}\Big(4,\frac{Q^2}{\mu^2} \Big)\,,
\eea
where ${\cal C}_{k,i}$ is the light parton coefficient function appearing in
$F_k(3,x,Q^2)$. The quantity which absorps the charm quark mass is given
by the operator matrix element $A_{ji}$ ($i,j=q,g$) which is the expectation
value of the heavy quark (here charm) operator sandwiched between light
quark and gluon states. Similarly one can write for $L_{k,i}^{\rm ASYMP}$ 
\bea
\label{eqn:5}
{\cal C}_{k,i}\Big(3,\frac{Q^2}{\mu^2}, \frac{m_c^2}{\mu^2}\Big) 
+ L_{k,i}^{\rm ASYMP}\Big (3,\frac{Q^2}{m_c^2}\Big ) =
A_{ji,c}\Big(3,\frac{m_c^2}{\mu^2}\Big) \otimes
{\cal C}_{k,j}\Big(4,\frac{Q^2}{\mu^2}\Big)\,.
\eea
The operator matrix elements $A_{ji,c}$ emerge from sandwiching the light 
partonic operators between light quark and gluon states which contain charm 
loop contributions to the gluon self energy only. 
Notice that the obove relations also involve a redefinition of the strong 
coupling constant which on the righthand side of the above equations depend
on four instead of on three flavours. For the more precise relations we refer
to \cite{bmsn1}. The mass factorization relations in Eqs. (\ref{eqn:4}),
(\ref{eqn:5}) can be also used when on the lefthand side the asymptotic
heavy quark functions are replaced by the exact ones. This has been done
by the authors in \cite{tr} to obtain modified coefficient functions 
${\cal C}_{k,i}$ on the righthand side.
which now become dependent on the charm mass. In this way one can preserve
the threshold behaviour of $F_{k,c}$ in their version of the variable flavour 
number scheme (VFNS) which differs from ours given below. Notice that using the
method in \cite{tr} there is an arbitrariness in fixing the mass dependent 
${\cal C}_{k,i}$.
After substituting the mass factorization relations in the sum 
$F_k(3,x,Q^2)+F_{k,c}(x,Q^2,m_c^2)$ one obtains $F_k(4,x,Q^2)$ which now
depend on four flavour parton densities including the one representing
the charm quark. The latter can be expressed into the original light parton
densities u,d,s and g as follows
\bea
\label{eqn:6}
f_{c+\bar c}(4, \mu^2) &\equiv&  f_{4}(4, \mu^2)
+ f_{\overline {4}}(4, \mu^2)
\nonumber \\
&=&  A_{cq}^{\rm S}\Big(3, \frac{m_c^2}{\mu^2}\Big)\otimes
\Sigma(3, \mu^2)
+  A_{cg}^{\rm S}\Big(3, \frac{m_c^2}{\mu^2}\Big) \otimes
G(3, \mu^2)\,,
\eea
with similar expressions for the light parton densities \cite{bmsn1}.
The above charm quark density has the property that it does not vanish
at $\mu=m_c$ for NLO in the ${\overline {MS}}$ scheme. 
Further one can construct a new charm component of the
structure function which is expressed into convolutions of the four light
quark densities, including charm, and the gluon density with the light
parton coefficient functions taken at four light flavours. This structure
function, presented in the four flavour number scheme (FFNS), is given by
\bea
\label{eqn:7}
&& F_{k,c}^{\rm PDF} (4,x, Q^2)   =   e_c^2 
\Big[f_{c +\bar c}\Big(4,\mu^2\Big) \otimes
{\cal C}_{k,q}^{\rm NS}\Big(4,\frac{Q^2}{\mu^2}\Big)
\nonumber\\[1ex]
&& + \Sigma\Big(4,\mu^2\Big) \otimes \tilde{\cal C}_{k,q}^{\rm PS}
\Big(4,\frac{Q^2}{\mu^2}\Big)
 + G\Big(4,\mu^2\Big) \otimes \tilde{\cal C}_{k,g}^{\rm S}
\Big(4,z,\frac{Q^2}{\mu^2}\Big) \Big]\,.
\eea
where $\tilde{\cal C}_{k,q}^{\rm PS}$ (purely singlet) is defined by
$\tilde{\cal C}_{k,q}^{\rm S}=\tilde{\cal C}_{k,q}^{\rm NS}+
\tilde{\cal C}_{k,q}^{\rm PS}$. The above expression stands for the
resummation of all large logarithms appearing in $F_{k,c}^{\rm ASYMP}$ 
and describes the charm structure function far above threshold which is
equivalent with large $Q^2$ and small $x$. An analysis \cite{bmsn1}
of the structure function $F_{k,c}^{\rm EXACT}$ (\ref{eqn:1}) (TFNS) and 
$F_{k,c}^{\rm PDF}$ (\ref{eqn:7}) (FFNS) reveals that the 
former gives the best description of charm production in the threshold
region whereas far away from threshold it is better to use the latter
structure function. Further one also needs a scheme which merges the advantages
of these two pictures and provides us with a good description of $F_{k,c}$ in
the intermediate region. This is given by the so called variable flavour
number scheme (VFNS) for which in \cite {bmsn1} we propose.
\bea
\label{eqn:8}
F_{2,c}^{\rm VFNS}(x,Q^2,m^2) &=&
F_{2,c}^{\rm PDF}(4,x,Q^2)
+ F_{2,c}^{\rm EXACT}(3,x,Q^2,m^2) \nonumber \\ &&
- F_{2,c}^{\rm ASYMP}(3,x,Q^2,m^2)\,.
\eea
In lowest order the above expression coincides with the VFNS scheme proposed
in \cite{acot}. The properties of $F_{2,c}^{\rm VFNS}$ are discussed in
\cite{bmsn1} and a comparison with the HERA data is made in \cite{bmsn2}.
From this comparison we conclude that at this moment the data do
not allow us to discriminate between the various schemes. They all give a 
reasonable good description of the data at large as well as at small $Q^2$.

\end{document}